\documentclass[12pt,preprint]{aastex}

 \newcommand{\pref}{\protect\ref}

\shorttitle{\indent \def Quasi-periodic upflows}
\shortauthors{Tian et al.}

\begin{document}

\title{The Spectroscopic Signature of Quasi-periodic Upflows in Active Region Timeseries}

\author{Hui Tian\altaffilmark{1}, Scott W. McIntosh\altaffilmark{1}, Bart De Pontieu\altaffilmark{2}}
\altaffiltext{1}{High Altitude Observatory, National Center for Atmospheric Research, P.O. Box 3000, Boulder, CO 80307; htian@ucar.edu; mscott@ucar.edu}
\altaffiltext{2}{Lockheed Martin Solar and Astrophysics Laboratory, 3251 Hanover St., Org. ADBS, Bldg. 252, Palo Alto, CA  94304; bdp@lmsal.com}

\begin{abstract}
Quasi-periodic propagating disturbances are frequently observed in coronal intensity image sequences. These disturbances have historically been 
interpreted as being the signature of slow-mode magnetoacoustic waves propagating into the corona. The detailed analysis of Hinode EUV Imaging Spectrometer (EIS) timeseries observations of an active region (known to contain propagating disturbances) shows strongly correlated, quasi-periodic, oscillations in intensity, Doppler shift, and line width. No frequency doubling is visible in the latter. The enhancements in the moments of the line profile are generally accompanied by a faint, quasi-periodically occurring, excess emission at  $\sim$100~km/s in the blue wing of coronal emission lines. The correspondence of quasi-periodic excess wing emission and the moments of the line profile indicates that repetitive high-velocity upflows are responsible for the oscillatory behavior observed. Furthermore, we show that the same quasi-periodic upflows can be directly identified in a simultaneous image sequence obtained by the Hinode X-Ray Telescope (XRT). These results are consistent with the recent assertion of De Pontieu \& McIntosh (2010) that the wave interpretation of the data is not unique. Indeed, given that several instances are seen to propagate along the direction of the EIS slit that clearly show in-phase, quasi-periodic variations of intensity, velocity, width (without frequency doubling), and blue wing enhanced emission this dataset would appear to provide a compelling example that upflows are more likely to be the main cause of the quasi-periodicities  observed here, as such correspondences are hard to reconcile in the wave paradigm.  
\end{abstract}


\section{Introduction}
Intensity oscillations with a period of three to thirty minutes have been frequently observed in polar plumes \citep[e.g.,][]{Ofman1997,DeForest1998,Ofman1999,Banerjee2010} and active region loops \citep[e.g.,][]{Berghmans1999,DeMoortel2000,DeMoortel2002,Robbrecht2001,Marsh2003,King2003,McEwan2006,Marsh2009,Stenborg2011}. In spectroscopic studies, small-amplitude oscillations (usually a few percent of the background emission) have been found in line intensities \citep{Banerjee2009} that are often accompanied by small fluctuations (at most a couple of km/s) in the Doppler velocities \citep{Wang2009a,Wang2009b,Kitagawa2010,Mariska2010}. These quasi-periodic disturbances usually show propagating speeds of 50-200~km/s and are almost interpreted as slow-mode magneto-acoustic waves propagating into the corona along the magnetic field without exception.

Recently, both imaging and spectroscopic observations have revealed that upflows with velocities of 50-150~km/s are prevalent in active regions \citep{Sakao2007,Hara2008,DePontieu2009,McIntosh2009a,He2010,Guo2010,Peter2010,Bryans2010}, quiet Sun \citep{McIntosh2009b} and coronal holes \citep{DePontieu2009,McIntosh2010a,McIntosh2010b}. These upflows appear as weak upward propagating disturbances in coronal images, and in spectroscopic observations they are identified as significant blue-wing asymmetries in emission profiles of spectral lines formed at transition-region and coronal temperatures. These faint upflows, with a life time of 50-150 seconds, are believed to be associated with type-II spicules or rapid blue-shifted events observed in the chromosphere \citep{DePontieu2009,Rouppe2009}. They are suggested to provide hot plasmas into the corona and may thus play an important role in coronal heating process \citep{DePontieu2009,McIntosh2009b,Peter2010,DePontieu2010,Hansteen2010}. 

These rapid upflows often recur at the same location on time scales of three to fifteen minutes \citep[e.g.,][]{McIntosh2009a, McIntosh2009b}, and would naturally cause quasi-periodic low-contrast oscillations in coronal images \citep[also see][]{Xia2005}. Thus, the discovery of these rapid quasi-periodic upflows challenges the universal wave interpretation of coronal oscillations. \cite{DePontieu2010} analyzed time series data that were previously studied by \cite{Wang2009b} to illustrate the presence of slow-mode waves, and found that the intensity and Doppler shift oscillations are {\em also} accompanied by oscillations in the line width and excess emission of the blue wing. They concluded that while the ``flows vs waves'' picture was not unambiguously resolved, the presence of these multi-moment in-phase oscillatory signatures is consistent with propagating quasi-periodic upflows causing the observed signature. 

In this paper, we present new results derived from a timeseries data set obtained by the EUV Imaging Spectrometer \citep[EIS,][]{Culhane2007} onboard Hinode. We show that coherent oscillatory behaviors in intensity, Doppler shift, line width, and blueward asymmetry are clearly present almost everywhere at the root of fan-like structures in the boundary of an active region. We demonstrate that these oscillation signatures are caused by repetitive high-speed upflows, which can be directly identified in the image sequence simultaneously obtained by the X-Ray Telescope \citep[XRT,][]{Golub2007} onboard Hinode. 

\section{Data reduction and analysis}

The EIS sit-and-stare data used here was acquired in AR 10942 from 17:50 to 20:45 on 2007 February 20.  The $1^{\prime\prime}\times512^{\prime\prime}$ slit was used for the observation, with a 30~s exposure and 32~s cadence. After standard correction and calibration of the EIS data, a running average over 3 pixels along the slit was applied to the spectra to improve the signal to noise ratio. We selected two strong emission lines in the spectral window for our study: Fe~{\sc{xii}}~195.12\AA{} and Fe~{\sc{xiii}}~202.04\AA{}. The Fe~{\sc{xii}}~195.12\AA{} line is known to be blended with the line Fe~{\sc{xii}}~195.18\AA{} line that is a few percent of the 195.12\AA{} brightness and, since it belongs to the same ion, it {\em should} exhibit the same Doppler behavior at approximately 100km/s in the red wing of the line observed by EIS \citep{Young2009}. It is common practice that this weak blend is ignored \citep[e.g.,][]{Harra2008,Wang2009b,Kitagawa2010,Tian2008,Tian2010} but this may not always be a safe practice \-- in regions where hot loop material is present the line can be significantly impacted by the wings of \ion{Fe}{14} (195.246\AA{}) or an unidentified line \citep[see, Table~2 of][]{Brown2008}; therefore great caution must be taken when using this line, especially when considering possible profile asymmetry analysis of the type introduced by \citet{DePontieu2009}. Again, as is common practice, we applied a single Gaussian fit to each EIS spectrum to derive the line intensity, Doppler shift and line width for each emission line. Using the method of \cite{Kitagawa2010}, we performed a cross-correlation analysis between intensities of two exposures to obtain the shift of the spectra, and thus to remove the jitter in the y-direction. In this case the jitter in the x-direction is neglected as it is comparable to the slit width. 

\begin{figure}
\centering {\includegraphics[width=0.4\textwidth]{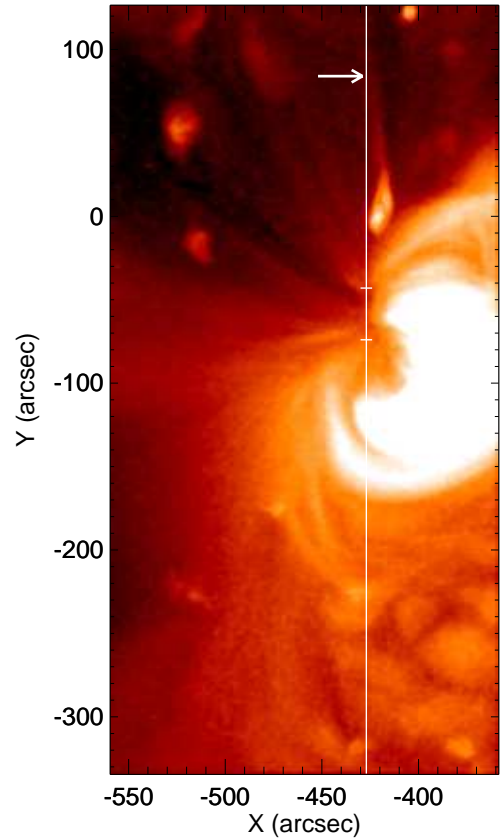}}
\caption{An XRT image of the studied active region at 19:40 showing the fan-like structure at its boundary. The white vertical line represents the location of the EIS slit. The section between the two horizontal bars representing the root of the fan is selected for further analysis. The arrow points to a strong isolated upflow event that travels across the EIS slit, see, Fig~\pref{fig.2}. The online edition of the journal has a animation of this figure.}
\label{fig.1}
\end{figure}

\begin{figure*}
\centering {\includegraphics[width=\textwidth]{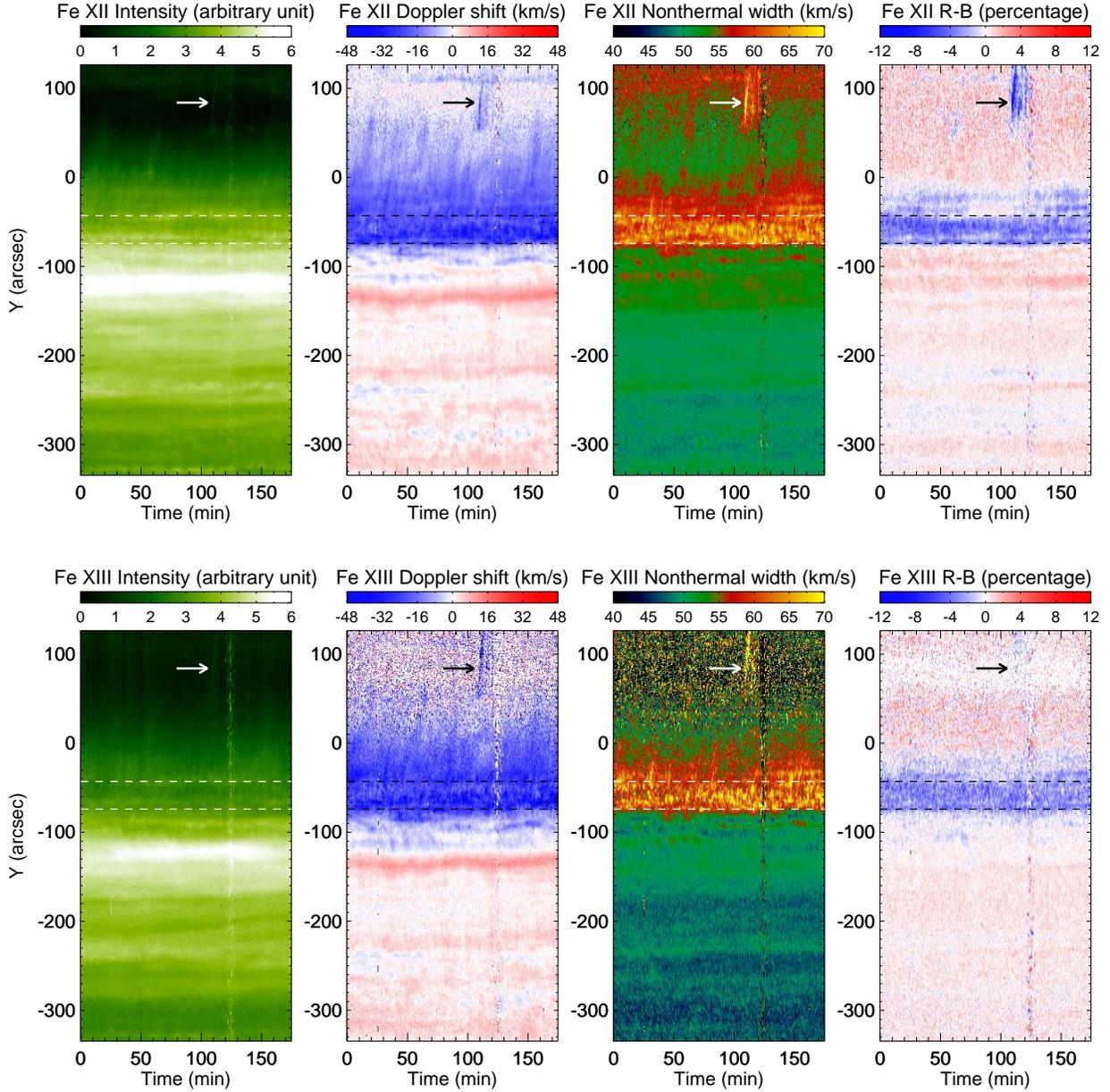}}
\caption{Temporal evolution of the line peak intensity, Doppler shift, Non-thermal width, and R-B asymmetry  (normalized to the peak intensity at each pixel) of the Fe~{\sc{xii}}~195.12\AA{}  and Fe~{\sc{xiii}}~202.04\AA{} lines. For the calculation of the Doppler shift, the rest wavelength of each line was determined by setting the average Doppler shift as zero. For the R-B asymmetry, negative and positive values correspond to blueward and redward asymmetries, respectively. The arrow points to an isolated strong upflow event. The section between the two dashed lines representing the root of the fan is selected for further analysis. }
\label{fig.2}
\end{figure*}

Results of the single Gaussian fit and ``R-B'' (red-blue) line asymmetry analysis \citep[][]{DePontieu2009} of the Fe~{\sc{xii}}~195.12\AA{} and Fe~{\sc{xiii}}~202.04\AA{} timeseries are shown in Fig.~\pref{fig.2}. By assuming the average Doppler shift of each line is zero over the entire space-time domain, we calculated the relative Doppler shift. We also calculated the non-thermal width of each line profile under the assumption of ionization equilibrium (ion temperature equals the formation temperature of the emission line). To quantify the asymmetry of each line profile, we performed a R-B analysis: after performing the initial single Gaussian fit to measure the line centroid, we interpolated the line profile to a spectral resolution ten times greater than the original one, then we simply subtracted the blue wing emission integrated over a narrow spectral range from that at the same position and over the same range in the red wing. The range of integration is then sequentially stepped outward from the line center to build an R-B profile. The example presented in Fig.~\pref{fig.2} shows the average of the R-B asymmetry from 80-140~km/s (normalized to the peak intensity at each pixel). Negative and positive values indicate asymmetries in the blue and red wings, respectively. Data in the time range between 122 and 127 minutes are affected by the spacecraft passage through the South Atlantic Anomaly (SAA).

At the same time, XRT continuously observed the AR in the Ti-Poly filter with a varying exposure time from 2 to 16~s. We selected those frames with an exposure time larger than 10~s for direct identification of upflow events. In total there are 99 images, with a mean cadence of 106~s. These images were first coaligned using a cross-correlation technique and then interpolated into regular time intervals. The location of the EIS slit in the image was determined by cross-correlating the EIS Fe~{\sc{xiv}}~264.78\AA{} line intensity along the slit and the XRT intensity at different x-locations. The accuracy of the coalignment is about $1^{\prime\prime}$ in the y-direction. Figure~\ref{fig.1} shows an XRT image obtained at 19:40. The white vertical line represents the EIS slit location.

\begin{figure*}
\centering {\includegraphics[width=\textwidth]{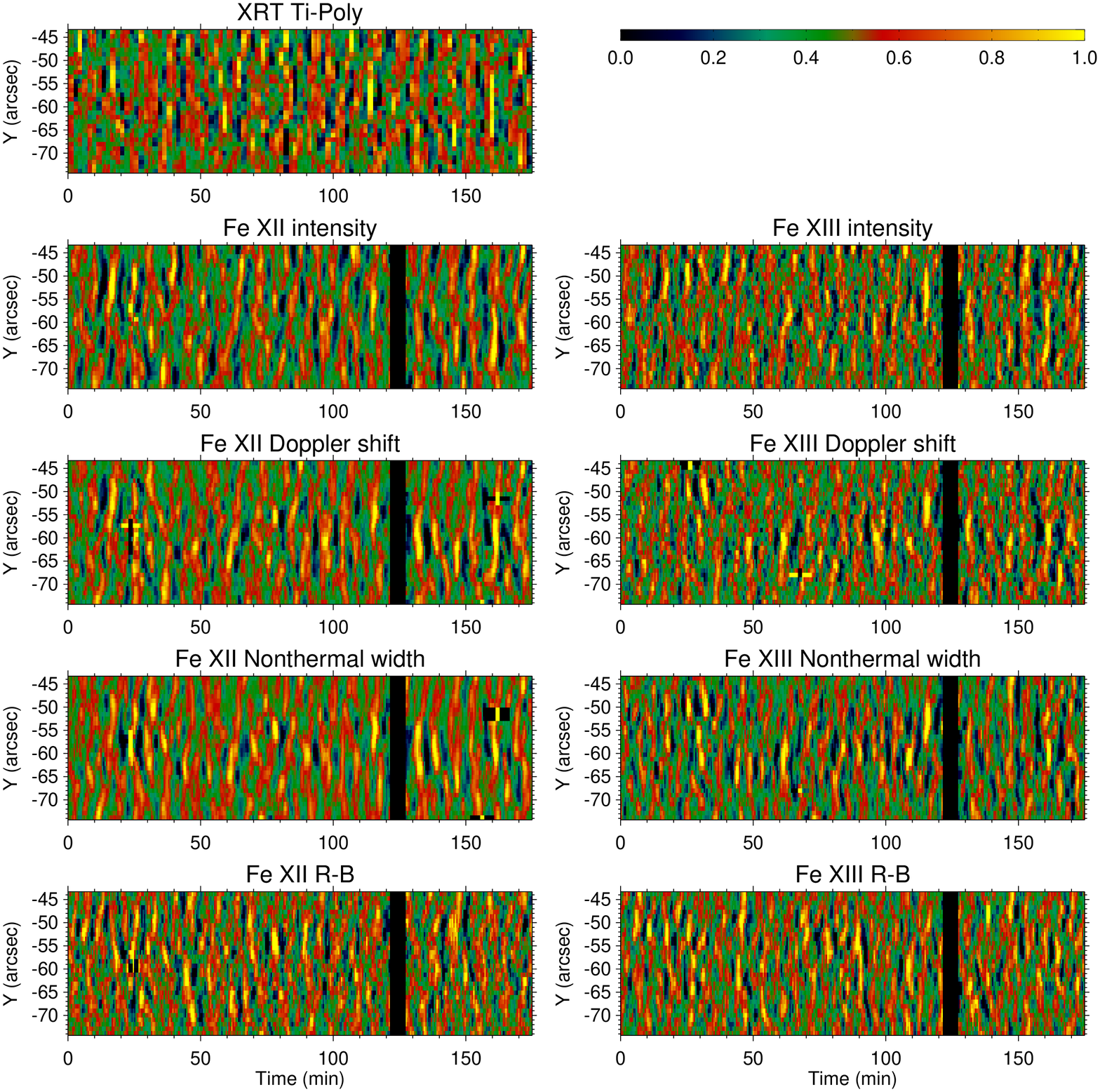}}
\caption{Temporal evolution of the de-trended XRT intensity and line parameters of Fe~{\sc{xii}}~195.12\AA{} and Fe~{\sc{xiii}}~202.04\AA{} in the section between the two horizontal bars in Figure~\ref{fig.1}. The values of Doppler shift and R-B are inverted so that a large value indicates more blueshifted emission or stronger blueward asymmetry. All parameters presented are de-trended in the way described in the text.}
\label{fig.3}
\end{figure*}

\begin{figure*}
\centering {\includegraphics[width=\textwidth]{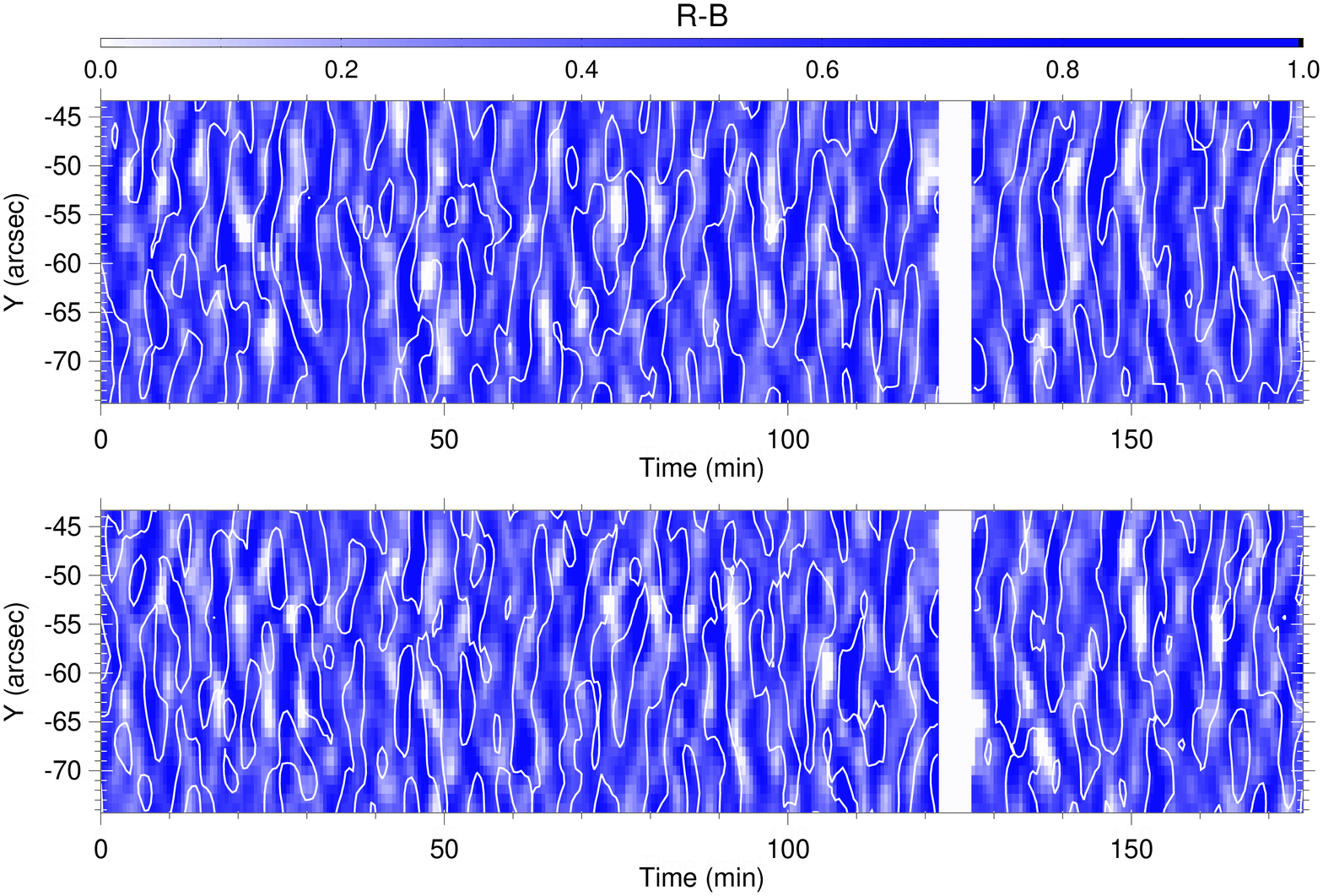}}
\caption{Contours of large non-thermal width (top 1/3) superposed on the R-B maps of the Fe~{\sc{xii}}~195.12\AA{} (upper) and Fe~{\sc{xiii}}~202.04\AA{} (lower) lines. Darker and lighter blue colors represent stronger and weaker blue asymmetries, respectively. }
\label{fig.4}
\end{figure*}

A comparison between Figure~\ref{fig.1} and Figure~\ref{fig.2} suggests that strong blue shift, large non-thermal broadening, and excess blueward emission are clearly present at the root of the fan-like structure in the boundary of the AR ($>$-83\arcsec). We selected the section between the two dashed lines in Figure~\ref{fig.2}, where the values of both Doppler shift and R-B are negative throughout almost the entire observation duration, for a more detailed analysis. The temporal evolution of the XRT intensity and line parameters of the Fe~{\sc{xii}}~195.12\AA{} and Fe~{\sc{xiii}}~202.04\AA{} lines in this section are presented in Figure~\ref{fig.3}. Here the parameters are de-trended at each slit location, this is done by subtracting a ten minute running average of each parameter from the timeseries. The de-trended intensities and R-B are then normalized to the local intensity. The values of Doppler shift and R-B are inverted such that a large positive value indicates more blue-shifted emission, or a stronger blueward asymmetry. In Figure~\ref{fig.4} we plot the contours of large non-thermal width on the R-B map of  the two lines. The data between 122 and 127 minutes were influenced by SAA and thus have been removed from the plot. 

\begin{figure*}
\centering {\includegraphics[width=0.9\textwidth]{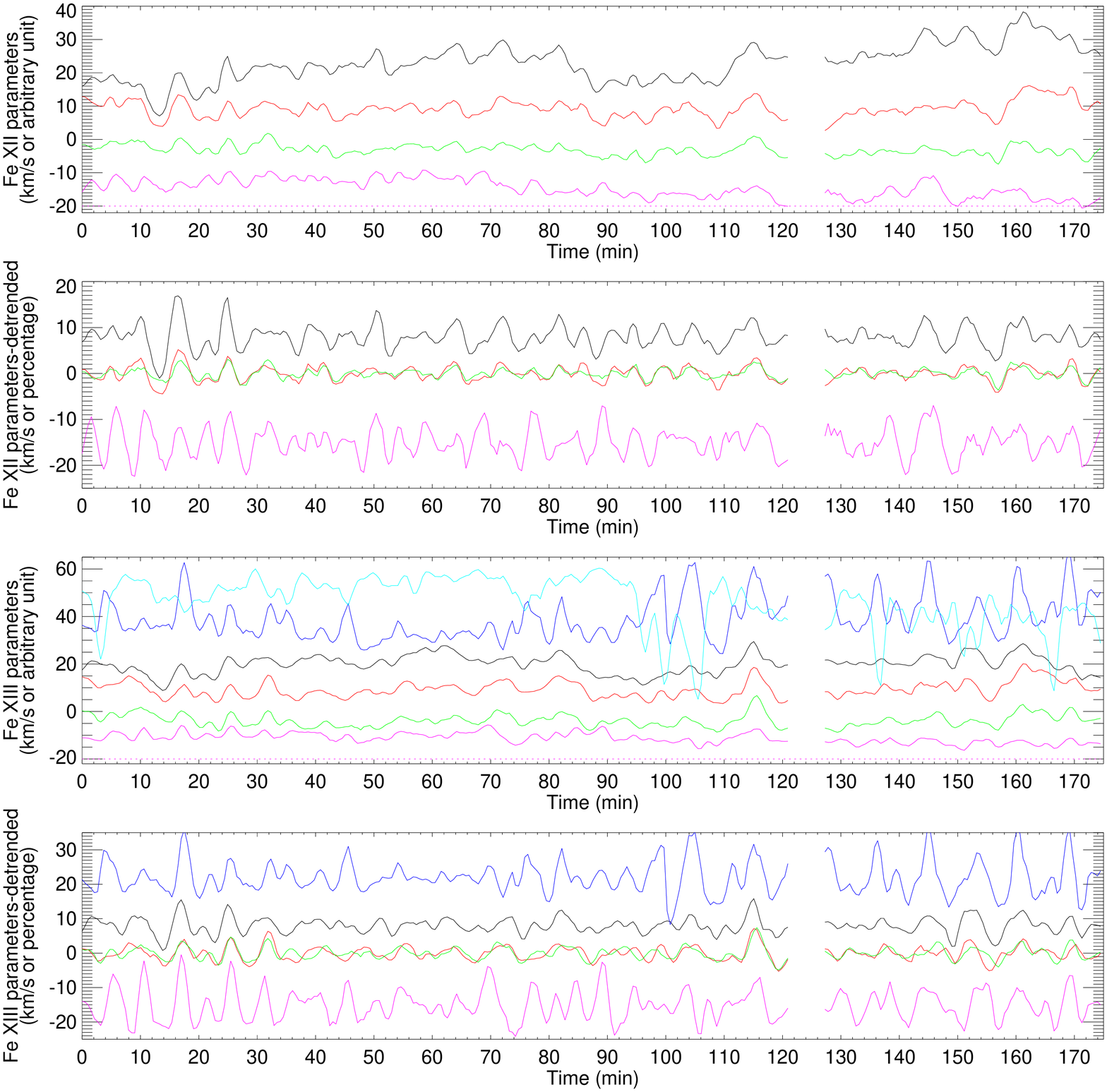}}
\caption{Timeseries for Fe~{\sc{xii}}~195.12\AA{} and Fe~{\sc{xiii}}~202.04\AA{}, averaged from y=-54$^{\prime\prime}$ to -50$^{\prime\prime}$. The black, red, green, violet, blue, and cyan curves represent the line intensity, Doppler shift, non-thermal width, R-B, intensity ratio between the secondary and primary Gaussian components, and the velocity difference of the two components, respectively. Note that the values of Doppler shift and R-B are inverted so that a large value indicates more blue-shifted emission or stronger blueward asymmetry. For the non-detrended parameters, the Doppler shift, non-thermal width, and velocity difference are shown in km/s, while other parameters are shown in arbitrary unit. The non-thermal width is offset by -65 on the y-axis, and the velocity difference is divided by 2, for the purpose of illustration within one frame. The R-B is also shifted on the y-axis and its zero line is drawn at y=-20, so that values above this line indicate blueward asymmetries. For the de-trended parameters, the intensity, R-B, and intensity ratio are shown in relative amplitude (percentage), while the Doppler shift and non-thermal width are shown in km/s. The intensity, R-B, and intensity ratio are offset by 8, -15, and 22 respectively on the y-axis. }
\label{fig.5}
\end{figure*}

We also applied a ``R-B guided'' double Gaussian fit to the EIS spectra with a high signal to noise ratio and a strong blueward asymmetry \citep[see, e.g., Sect.~5 of][]{DePontieu2010}. We used the centroid of the R-B asymmetry profile as an initial guess of the spectral position of the secondary component. It turns out that a blue-shifted secondary component is clearly present in the region where strong blueward profile asymmetry is found. An example of timeseries for both Fe~{\sc{xii}}~195.12\AA{} and Fe~{\sc{xiii}}~202.04\AA{} is presented in Figure~\ref{fig.5}. The double Gaussian fit was not performed for the Fe~{\sc{xii}}~195.12\AA{} line since the weak blend at the red wing could have an important impact on the result produced by the sensitive fitting algorithm. For a better illustration, we have smoothed the parameters over three adjacent time steps. Again, the values of Doppler shift and R-B are inverted. Again, the data between 122 and 127 minutes have been removed from the plot. 

\section{Results and Discussion}
Applying a single Gaussian fit to the line profiles, it has been found that coronal emission lines usually show blue shift of the order of 30~km/s at boundaries of some ARs \citep{Marsch2004,Marsch2008,Harra2008,DelZanna2008,Doschek2008,Murray2010}. These blue shifts were thought to be genesis of the slow solar wind \citep{Sakao2007,Harra2008,Doschek2008}. However, from Figure~\ref{fig.2} we can see that the Fe~{\sc{xii}}~195.12\AA{} and Fe~{\sc{xiii}}~202.04\AA{} profiles at the root of the fan-like structure (y=-83$^{\prime\prime}$$\sim$-43$^{\prime\prime}$) in the AR boundary are actually very asymmetric with prominent excess emission in blue wings, confirming the results of \cite{McIntosh2009a}. Although the Fe~{\sc{xii}}~195.12\AA{} line is potentially blended, we place great confidence in its strong blue wing asymmetry here since the potential blends are sitting in the red wing of the profile. Such a result suggests the presence of continuous fast-moving upflows (around 100~km/s) and that the emission consists of multiple components. The centroid of the line profile derived by using a single Gaussian fit only reflects the ensemble velocity of the multiple emission components. And different components may have different velocities, which naturally broadens the profile. 

The lack of blue asymmetry above y=-43$^{\prime\prime}$ seems to suggest that rapid upflows occur mainly at the root of the fan-like structures. Perhaps the larger field inclination with height along the loops (thus larger angle between the flow direction and the line of sight) makes it difficult to resolve the upflow signatures. While the profiles of Fe~{\sc{xiii}}~202.04\AA{} are noisy there, the Fe~{\sc{xii}}~195.12\AA{} line profiles reveal clear prominent red wing asymmetries, which might be caused by the blends of the line (see above). The lower part of the slit (y$\leq$-83$^{\prime\prime}$) is dominated by loop structures in the AR core, where weak red wing asymmetries are present in profiles of both lines. It might be related to the complexity of the emission and magnetic structures. The asymmetries are more prominent for the Fe~{\sc{xii}}~195.12\AA{} line, which should be related to the blends.

Examining the XRT movie associated with Fig.~\pref{fig.1}, we can see plasma moving outward rapidly along loop structures in the fan. These continuous upflows have previously been identified by \cite{Sakao2007} having an average speed of $\sim$100~km/s. We believe that these rapid upflows are responsible for the blueward asymmetry in coronal emission line profiles, consistent with the results of \citet{McIntosh2009a}. An isolated strong upflow event (indicated by the arrow in Figure~\ref{fig.1} and Figure~\ref{fig.2}) was clearly visible in both the imaging and spectroscopic observations. As the rapidly moving plasma crossed the slit at around 19:40, we immediately observed a significant enhancement of the emission in the blue wing of the emission line. This enhanced blue wing asymmetry resulted in an enhancement in the line intensity, Doppler shift, and non-thermal width derived by a single Gaussian fit. This we, believe is an isolated example of the process occurring frequently at the roots of the fan structure and is highly unlikely to be the result of wave passage.

Although the cadence of the XRT observation is much lower than that of EIS, and the plasma sampled by the XRT Ti-Poly filter has a higher temperature than the formation temperature of Fe~{\sc{xii}}, from Figure~\ref{fig.3} we see a good correspondence in the temporal evolutions of the XRT and coronal line intensities. Figure~\ref{fig.3} shows that the evolutionary patterns of the de-trended intensity, Doppler shift, and non-thermal width are highly similar (the correlation coefficient between each pair of line parameters is around 0.6). The R-B pattern, to some extent, is also similar to those of the other four. In Figure~\ref{fig.4} we find that patches of large non-thermal width (contours) often coincide with those of large blue asymmetry (darker colors). The correlation coefficient between non-thermal width and R-B is 0.40 for Fe~{\sc{xii}}~195.12\AA{} and 0.29 for Fe~{\sc{xiii}}~202.04\AA{}. The relatively low value for the latter might be partly caused by the much lower signal to noise ratio of the Fe~{\sc{xiii}}~202.04\AA{} line profiles. The coherent behaviors revealed in Figure~\ref{fig.3} and Figure~\ref{fig.4} strongly suggest that continuous upflows with quasi-periodic enhancement of the flow intensity are responsible for the quasi-periodic enhancement of the line intensity, Doppler shift, and non-thermal width determined from a single Gaussian fit. We note that the correlations at several instances are not obvious or even not present. This is likely to be caused by the poor spectral resolution and high photon noise of the EIS instrument \citep{DePontieu2010}.

As an example, Figure~\ref{fig.5} shows the timeseries for Fe~{\sc{xii}}~195.12\AA{} and Fe~{\sc{xiii}}~202.04\AA{} at y=-54$^{\prime\prime}$$\sim$-50$^{\prime\prime}$. Correlated changes in intensity, Doppler shift, non-thermal line width, and R-B are clearly present. It is also clear that the intensity ratio between the secondary and primary Gaussian components generally varies with R-B, suggesting that the fast-moving plasma is resolved by our guided double Gaussian fit. Note that the less-than-ideal correlations at some instances are actually the result of the poor spectral resolution and photon noise of the EIS instrument \citep{DePontieu2010}. The relative velocity of the secondary component is rather stable at $\sim$ 100~km/s, except for several instants when the R-B values are relatively small. 

As we mentioned in the introduction, quasi-periodic intensity oscillations have been almost universally interpreted as waves. However, from Figures~\ref{fig.3}-\ref{fig.5}, it is clear that repetitive flows can also produce oscillatory signatures, as demonstrated previously by \cite{DePontieu2010}. Recently, \cite{Verwichte2010} presented a slow wave model to argue that the wave interpretation is still valid for the observed quasi-periodic intensity perturbations. However, the Figure~3 of their paper shows a frequency-doubling of the line width oscillation compared to the intensity and Doppler shift oscillations, which is {\em not} observed by EIS in our observation. Instead, all of these single Gaussian parameters show a reasonable correlation over several hours. Moreover, in the wave scenario the R-B values often can be positive and negative over an oscillation period \--- this is also {\em not} the case in our observation. From our Figures~\ref{fig.2} and~\ref{fig.5} we can see that in the fan root region the R-B values almost remain the same sign in the entire 175-minute observation period, indicating the presence of continuous blueward emission from upflows. 

In conclusion, we find that coherent oscillatory behaviors in intensity, Doppler shift, line width, and blueward asymmetry are clearly present almost everywhere at the root of the fan-like structures in the boundary of an active region. With coordinated imaging observation, we conclude that the quasi-periodicities we observed are more likely to be caused by quasi-periodic high-speed upflows. There is no doubt that both waves and flows are present on the Sun. We emphasize that it is difficult to distinguish between upflows and waves only through the intensity evolution. Spectroscopic observations reveal more information, and a combination of imaging and spectroscopic observations is critical for the correct interpretation. So far we have found two observations where flows seem to be a better interpretation for the quasi-periodicity. With more detailed spectroscopic observations we are sure that we can find more evidences of flows. 

\begin{acknowledgements}
EIS and XRT are instruments onboard {\it Hinode}, a Japanese mission developed and launched by ISAS/JAXA, with NAOJ as domestic partner and NASA and STFC (UK) as international partners. It is operated by these agencies in cooperation with ESA and NSC (Norway). Hui Tian is supported by the ASP Postdoctoral Fellowship Program of National Center for Atmospheric Research. The work carried out in this manuscript was partly supported by NASA (NNX08AL22G, NNX08BA99G) and NSF (ATM-0541567, ATM-0925177). The National Center for Atmospheric Research is sponsored by the National Science Foundation. We thank Alfred de Wijn and the anonymous referee for helpful comments. 

\end{acknowledgements}

\end{document}